\documentclass[aps,pra,floatfix,twocolumn]{revtex4}
\usepackage[dvips]{graphicx}

\usepackage{longtable}
\usepackage{dcolumn}
\usepackage[dvips]{graphicx}
\usepackage{bm}
\usepackage{bbm}

\usepackage{times}
\usepackage{nicefrac}
\usepackage{amsmath}
\usepackage{amsfonts}
\usepackage{amssymb}
\usepackage{amsthm}

\newcolumntype{.}{D{x}{}{-1}}

\usepackage{color}

\newcommand{\vare}{\varepsilon}

\newcommand{\Za}{Z\alpha}

\begin{document}

\title{Lamb shift of $\bm n \bm = \bm 1$ and $\bm n \bm = \bm 2$ states of hydrogenlike atoms, $\bm{ 1 \le Z \le 110}$}

\author{V. A. Yerokhin} \affiliation{Center for Advanced Studies,
        Peter the Great St.~Petersburg Polytechnic University, Polytekhnicheskaya 29,
        St.~Petersburg 195251, Russia}
\author{V. M. Shabaev} \affiliation{Department of Physics, St.~Petersburg State University, Ulianovskaya 1, Petrodvorets,
St.Petersburg 198504, Russia}

\begin{abstract}
 Theoretical energy levels of the $n = 1$ and $n = 2$ states of hydrogenlike atoms with the
 nuclear charge numbers $1 \le Z \le 110$ are tabulated. The tabulation is based on {\em ab
 initio} quantum electrodynamical calculations performed to all orders in the nuclear binding
 strength parameter $\Za$, where $\alpha$ is the fine structure constant. Theoretical errors due
 to various effects are critically examined and estimated.
\end{abstract}


\maketitle

\section{Introduction}

Hydrogen is the simplest atom, with the nucleus surrounded by a single electron. A hydrogenlike
atom is a bound system isoelectronic with hydrogen, consisting of any atomic nucleus and an
electron. Except for the hydrogen itself, hydrogenlike atoms carry a positive charge $|e|(Z-1)$,
where $e$ is the electron charge and $Z$ is the nuclear charge number of the atom. Examples of
hydrogenlike atoms are He$^+$, Li$^{2+}$, U$^{91+}$.


The simplicity of hydrogenlike atoms makes them an ideal testing ground for extending the theory
based on the principles of quantum electrodynamics (QED) up to the utmost precision \cite{mohr:98}.
The creation of the modern QED goes back to the late 1940s, when Lamb and Retherford discovered the
$2s$-$2p_{1/2}$ splitting in hydrogen, which is now known as the Lamb shift. The first calculation
of the Lamb shift was performed by Hans Bethe \cite{bethe:47} who used Kramer's idea of the mass
renormalization. The rigorous formalism of QED was developed by Dyson, Feynman, Schwinger and
Tomonaga, who proved that all infinities in the theory can be removed by the so-called
renormalization procedure.

All calculations in QED are based on the perturbation expansion in powers of the fine structure
constant $\alpha \approx 1/137.036$. The individual terms of the perturbation series can be
conveniently represented in terms of the so-called Feynman diagrams, with the power of $\alpha$
corresponding to the number of virtual-photon exchanges.

Before the 1970s, investigations of QED effects in atomic systems were mainly focused on low-$Z$
atoms such as hydrogen and helium. In these systems, in addition to the expansion parameter
$\alpha$, there is another small parameter $\Za$, which characterizes the strength of the nuclear
binding field. Because of this, most early calculations of QED effects were based on the expansion
in both $\alpha$ and $\Za$. Such calculations produced accurate results for hydrogen and a few
light atoms but provided little or no information about QED effects in high-$Z$ ions, in which the
parameter $\Za$ approaches unity.

In 1974 P.~J.~Mohr reported his pioneering calculations of the electron self-energy to all orders
in the parameter $\Za$ \cite{mohr:74:a,mohr:74:b}. These calculations, together with their
extension to the $n = 2$ states several years later \cite{mohr:82}, have greatly increased the
theoretical accuracy of the Lamb shift in hydrogenlike ions. The results of these studies were used
as a basis of the compilations by Mohr \cite{mohr:83} and by Johnson and Soff \cite{johnson:85},
which for many years remained the standard references on hydrogenlike ions.

During the decades following the compilation by Johnson and Soff, many calculations of various
effects contributing to the Lamb shift of hydrogenlike atoms were performed. Most notably, the
vacuum-polarization was calculated to all orders in $\Za$ in
Refs.~\cite{soff:88:vp,manakov:89:zhetp}, the nuclear size correction to the self-energy was
obtained in Ref.~\cite{mohr:93:prl}, the nuclear recoil correction was evaluated to the first order
in $m/M$ but to all orders in $\Za$ in Ref.~\cite{artemyev:95:pra}, the nuclear polarization
correction was obtained in Ref.~\cite{plunien:95}, the two-loop self-energy was calculated to all
orders in $\Za$ in Refs.~\cite{yerokhin:01:sese,yerokhin:03:prl}. Alongside with the above
mentioned nonperturbative calculations, the hydrogen theory was developed further within the
$\Za$-expansion approach. The most notable achievements were the calculation of the
$\alpha^2\,(\Za)^5$ two-loop contribution \cite{pachucki:94,eides:95:pra} and the calculations of
the two-loop corrections to order $\alpha^2\,(\Za)^6$
\cite{pachucki:01:pra,pachucki:03:prl,jentschura:05:sese}.

Theoretical advances of the last decades were complemented by the experimental progress in
high-precision spectroscopy of highly charged ions \cite{beyer:03}. Modern experiments access
hydrogenlike ions along the whole Periodic Table up to uranium and study the strong-nuclear-field
regime of the bound-state QED. Measurements of the ground-state Lamb shift in the hydrogenlike
uranium have rapidly progressed during last years, achieving the accuracy of
$4.6$~eV~\cite{gumberidze:05:prl}, which corresponds to the fractional error of 1\% of the Lamb
shift. Further advance is anticipated in these experiments, which will make them sensitive to the
nonperturbative two-loop QED effects.

The goal of the present work is to summarize theoretical developments of the last decades and to
compile an up-to-date tabulation of the energy levels of the $n = 1$ and $n = 2$ states of
hydrogenlike atoms with the nuclear charges $1 \le Z \le 110$.

\section{Theory}

The total energy of a hydrogenlike atom is a sum of the rest mass of the atom and the electron
binding energy $E_{\rm bind}$,
\begin{align} \label{eq1}
E &\ =  Mc^2 + mc^2 + E_{\rm bind}\,,
\end{align}
where $M$ is the nuclear mass, $m$ is the electron mass, and $c$ is the speed of light.

The dominant part of the electron binding energy of a hydrogenlike atom can be calculated
analytically. Several closed-form analytical results are well-known, namely: (i) the Dirac equation
for the point-nucleus Coulomb potential can be solved exactly to all orders in $\Za$, (ii) the
nonrelativistic two-body problem for the point particles can be solved for arbitrary masses $M$ and
$m$, (iii) the recoil correction within the Breit approximation can be obtained analytically to
first order in $m/M$ but to all orders in $\Za$. It is natural to separate out these analytical
contributions to the binding energy, with the rest being called as the Lamb shift. So, we write the
electron binding energy of a hydrogenlike atom as
\begin{align} \label{eq1a}
E_{\rm bind} &\ =  E_{\rm D} + E_{\rm M} + E_{\rm L}\,,
\end{align}
where $E_{\rm D}$ is the Dirac point-nucleus biding energy, $E_{\rm M}$ is the leading nuclear mass
(recoil) correction, and $E_{\rm L}$ is the Lamb shift.

The Dirac point-nucleus binding energy $E_{\rm D}$ is given by
\begin{align} \label{eq2}
E_{\rm D} = mc^2\, \left[
\sqrt{1 - \frac{(\Za)^2}{N^2}} - 1
\right]\,,
\end{align}
where
\begin{align} \label{eq:N}
N &\ = \sqrt{(n_r+\gamma)^2 + (\Za)^2}\,,\\
\gamma &\ = \sqrt{\kappa^2-(\Za)^2}\,,
\end{align}
$n_r = n - |\kappa|$ is the radial quantum number, $n$ is the principal quantum number, and
$\kappa$ is the angular momentum-parity quantum number.

The leading recoil correction $E_{\rm M}$ is written as
\begin{align} \label{eq3}
E_{\rm M} =&\ \frac{m^2c^4 - \vare_D^2}{2Mc^2} - \left[ m_r - m + \frac{m^2}{M}\right]c^2\,\frac{(\Za)^2}{2n^2}
 \nonumber \\
=&\  mc^2\,\frac{m}{M}\frac{(\Za)^2}{2N^2} - m_r c^2 \,\left( \frac{m}{M}\right)^2 \,
   \frac{(\Za)^2}{2n^2}\,,
\end{align}
where $\vare_D = mc^2+ E_D$ and $m_r = mM/(m+M)$ is the reduced mass of the electron. $E_M$
consists of two parts. The first part is the first-order (in $m/M$) recoil correction as derived
from the Breit equation \cite{shabaev:85}, whereas the second part is the nonrelativistic
reduced-mass correction of second and higher orders in $m/M$.

$E_{\rm L}$ is the remaining part of Eq.~(\ref{eq1a}), which is the definition of the Lamb shift
adopted in the present work. We note that the separation (\ref{eq1a}) is not unique and different
definitions of the Lamb shift can be found in the literature. In the high-$Z$ region, the recoil
effect is a small correction, so it is often included into the Lamb shift
\cite{mohr:98,shabaev:02:rep}. For low-$Z$ ions, however, the nuclear recoil is a large effect, so
its leading part is usually separated out. In the theory of hydrogen it is common
\cite{mohr:12:codata} to define $E_{\rm M}$ in a slightly different way, which is based on the
result derived from the modified Dirac equation \cite{sapirstein:90:kin}. The difference with the
present definition is very small and is noticeable for the case of hydrogen and He$^+$ only. We
prefer to define $E_{\rm M}$ in the form of Eq.~(\ref{eq3}) since it does not contain any
contributions of order $(m/M)^2(\Za)^4$ (having in mind that the full correction to this order
depends on the nuclear spin) or any terms of higher orders that are not known at present. We also
note a small difference of definitions of the Lamb shift in the present compilation and in the
tabulation by Johnson and Soff \cite{johnson:85}, which is of no importance at the level of
accuracy of Ref.~\cite{johnson:85}.

\subsection{Lamb shift}

\subsubsection{One-loop QED correction}

The dominant contribution to the Lamb shift comes from the one-loop QED effects, namely, the
electron self-energy and the vacuum-polarization.

The {\em one-loop self-energy correction} for the point nuclear charge is represented as
\begin{align} \label{se:1}
E_{\rm SE, pnt} = mc^2\,\left(\frac{m_r}{m} \right)^3\, \frac{\alpha}{\pi}\, \frac{(\Za)^4}{n^3}\,
  F_{\rm SE, pnt}(\Za)\,,
\end{align}
where
\begin{align} \label{eq:se:za}
F_{\rm SE, pnt}(\Za) &\ = L\,A_{41} + A_{40} + (\Za)\,A_{50}
  \nonumber \\ &
+ (\Za)^2\,\biggl[L^2\,A_{62} + L\,A_{61}+ G_{\rm SE,pnt}(\Za)\biggr]\,,
\end{align}
$L = \ln\left[(m/m_r)(\Za)^{-2}\right]$, and  $G_{\rm SE, pnt}(\Za) = A_{60} + \ldots$ is the
remainder that contains all higher-order expansion terms in $\Za$.

The first coefficients $A_{ij}$ of the $\Za$ expansion are well known (see Ref.~\cite{erickson:65}
and earlier papers cited therein),
\begin{align}
A_{41} &= \frac43\,\delta_{l,0}\,,\\
A_{40} &= -\frac43\,\ln k_0(n,l) + \frac{10}{9}\,\delta_{l,0} - \frac{m/m_r}{2\kappa(2l+1)}
(1-\delta_{l,0})\,, \\
A_{50} &= \left( \frac{139}{32} - 2 \ln 2\right)\,\pi \,\delta_{l,0}\,,\\
A_{62} &= -\delta_{l,0}\,,
\end{align}
where $\ln k_0(n,l)$ is the Bethe logarithm, whose numerical values for the states of the present
interest are \cite{drake:90}
\begin{align}
\ln k_0(1s) & = 2.984\,128\,556\,,\\
\ln k_0(2s) & = 2.811\,769\,893\,,\\
\ln k_0(2p) & =-0.030\,016\,709\,.
\end{align}
The numerical values of the $A_{61}$ coefficient for the $n = 1$ and $n = 2$ states are (see
Ref.~\cite{erickson:65} for analytical formulas)
\begin{align}
A_{61}(1s) &= 5.419\,373\,685\,,\\
A_{61}(2s) &= 5.930\,118\,296\,,\\
A_{61}(2p_{1/2}) &= 0.572\,222\,222\,,\\
A_{61}(2p_{3/2}) &= 0.322\,222\,222\,.
\end{align}

The higher-order remainder $G_{\rm SE,pnt}$ is obtained from numerical computations performed to
all orders in $\Za$. First accurate numerical calculations of the electron self-energy were
performed by P.~J.~Mohr for the point nuclear charge and ions with $Z \ge 10$ for the $1s$ state in
Refs.~\cite{mohr:74:a,mohr:74:b} and for the $n = 2$ states in Ref.~\cite{mohr:82}. Updated results
with improved precision were reported for $Z \ge 5$ and $n = 1$ and $n = 2$  in
Ref.~\cite{mohr:92:b}. Accurate nonperturbative calculations of the electron self-energy for the
lightest hydrogenlike atoms with $Z \leq 5$ were presented by Jentschura, Mohr and Soff in
Refs.~\cite{jentschura:99:prl,jentschura:01:pra}. Different methods for the nonpertubative
calculation of the self-energy correction were developed in
Refs.~\cite{snyderman:91,blundell:91:se,indelicato:92:se,persson:93:ps,quiney:93,cheng:93,yerokhin:99:pra,yerokhin:05:se}.

In the present compilation, we used the numerical results from Ref.~\cite{jentschura:01:pra} for $Z
\leq 5$ and from Ref.~\cite{yerokhin:05:se} for $5 < Z \leq 10$. For $Z = 15$, $20$, $25$, $26$,
$30$, $35$, $36$, $40$, $45$, $50$, $54$, $55$, $60$, $65$, $66$, $70$, $75$, $79$, $80$, $82$,$
83$, $85$, $90$, $92$, $95$, $100$, $105$, and $ 110$ we used the numerical results from
Ref.~\cite{mohr:92:b}. For other $Z$, we performed our own calculations by the method described in
Refs.~\cite{yerokhin:99:pra,yerokhin:05:se}. Individual calculations were carried out for each
value of $Z$, so that no loss of precision due to interpolation occurred. Our results were checked
to be fully consistent with those of Ref.~\cite{mohr:92:b} within the estimated numerical errors.

The {\em vacuum-polarization} is the QED shift of the energy levels due to production of a virtual
$e^+e^-$ pair. The dominant part of the one-loop vacuum polarization is known as the Uehling
correction~\cite{uehling:35}. The Uehling approximation treats the virtual $e^+e^-$ pair as freely
propagating, i.e., undisturbed by the nuclear field. This treatment takes into account the virtual
$e^+e^-$ pair to the leading order in the coupling constant $\Za$.  The remaining part of the
one-loop vacuum-polarization is referred to as the Wichmann-Kroll correction \cite{wichmann:56}.
The Wichmann-Kroll potential contains all higher-order terms in $\Za$, $\sim (\Za)^n$ with $n \ge
3$. According to the Furry theorem, there are no contributions $\sim (\Za)^n$ with even powers of
$n$.

The one-loop Uehling vacuum-polarization correction for the point nuclear charge is represented as
\cite{erickson:65}
\begin{align} \label{ueh:1}
E_{\rm Ueh, pnt} = mc^2\,\left(\frac{m_r}{m} \right)^3\, \frac{\alpha}{\pi}\, \frac{(\Za)^4}{n^3}\,
  F_{\rm Ueh, pnt}(\Za)\,,
\end{align}
where
\begin{align} \label{eq:ueh:za}
F_{\rm Ueh, pnt}(\Za) = &\ -\frac{4}{15}\,\delta_{l,0} + \frac{5}{48}\pi(\Za)\,\delta_{l,0}
  \nonumber \\ &
-\frac{2}{15} (\Za)^2\, L\, \delta_{l,0}
  \nonumber \\ &
+ (\Za)^2 \, G_{\rm Ueh,pnt}(\Za)\,,
\end{align}
and $G_{\rm Ueh,pnt}$ is the higher-order remainder. The calculation of $G_{\rm Ueh,pnt}$  is quite
straightforward and can be performed up to arbitrary accuracy. The results listed in the present
compilation are obtained by our own calculations; they fully agree with other values published
previously in the literature, e.g., in Ref.~\cite{mohr:82}.

The {\em one-loop Wichmann-Kroll vacuum-polarization correction} for the point nuclear charge is
parameterized as
\begin{align}\label{wk:1}
E_{\rm WK, pnt} = mc^2\,\left(\frac{m_r}{m} \right)^3\, \frac{\alpha}{\pi}\, \frac{(\Za)^6}{n^3}\,
  G_{\rm WK, pnt}(\Za)\,.
\end{align}
Unlike the Uehling contribution, a direct calculation of the Wichmann-Kroll correction is a
non-trivial task. First calculations of this correction to all orders in $\Za$ were reported by
Soff and Mohr for the shell model of the nuclear charge distribution \cite{soff:88:vp} and by
Manakov and co-workers for the point nuclear charge \cite{manakov:89:zhetp}. Accurate calculations
of the Wichmann-Kroll correction were later performed also by other groups
\cite{persson:93:vp,sapirstein:03:vp,yerokhin:11:fns}. Recently, the problem of calculating the
Wichmann-Kroll correction for the point nuclear charge was greatly simplified by the high-precision
approximation formulas reported by Manakov and Nekipelov \cite{manakov:12:vgu,manakov:13:vgu}.
These formulas are based on the analytical expansions of Whittaker functions and approximate the
exact point-nucleus Wichmann-Kroll potential with a relative accuracy of $10^{-6}$ or better for
nuclear charges $Z \le 100$. Since this precision is fully sufficient in the context of the present
compilation, we used the formulas from Ref.~\cite{manakov:12:vgu} to calculate the Wichmann-Kroll
correction for the point nuclear charge. The results obtained in this way are in agreement with the
direct calculations of the Wichmann-Kroll correction for the point nuclear charge
\cite{manakov:89:zhetp,fainshtein:91}.

\begin{figure}
\centerline{
\resizebox{\columnwidth}{!}{%
  \includegraphics{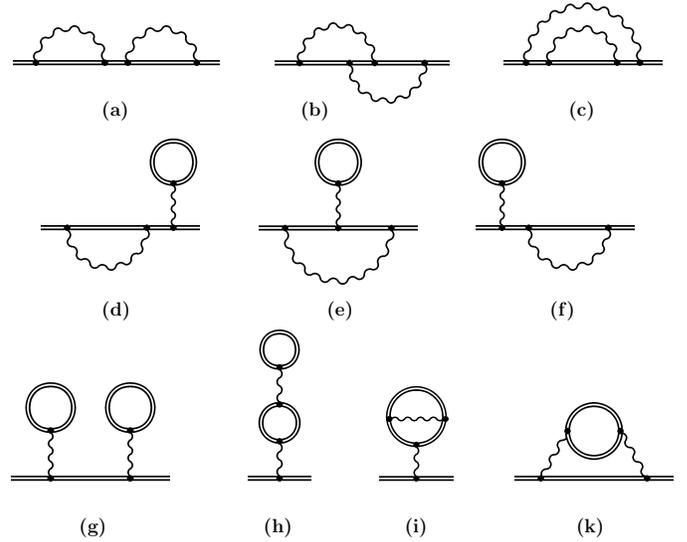}
}}
 \caption{Two-loop QED corrections. The double line represents the electron propagating in the binding nuclear field; the wavy line
 denotes the virtual photon.
\label{fig:2order}}
\end{figure}

\begin{table*}
\caption{Coefficients of the $Z\alpha$ expansion of the two-loop corrections. \label{tab:twoloop}}
\begin{ruledtabular}
\begin{tabular}{llccccc}
     &      &  $B_{40}$ & $B_{50}$ & $B_{63}$ & $B_{62}$ & $B_{61}$\\ \hline \\[-2pt]
$1s$ & SESE & $1.409244$ & $-24.26506(13)$ & $-8/27$   & $-0.639669$  &  $48.388913$ \\
     & SEVP & $0.142043$ & $1.305370$ &  $0$      &  $8/45$     & $1.111406$  \\
     & VPVP & $-164/162$ & $1.405241$ &  $0$      &  $0$         & $-1097/2025$ \\[5pt]
$2s$ & SESE & $1.409244$ & $-24.26506(13)$ & $-8/27$   &   $0.461403$ & $40.932915$ \\
     & SEVP & $0.142043$ & $1.305370$ &  $0$        &    $8/45$   & $0.670977$ \\
     & VPVP & $-164/162$ & $1.405241$ &  $0$      &  $0$         & $-1097/2025$ \\[5pt]
$2p_{1/2}$
     & SESE & $0.114722\,(m/m_r)$ &  $ $0$   $ &  $0$    &   $1/9$      & $0.202220$ \\
     & SEVP & $-0.005229\,(m/m_r)$ & $ $0$   $ &  $0$    &  $0$        & $-2/45$ \\
     & VPVP & $0       $ & $ $0$    $&  $0$    &  $0$        &$0$\\[5pt]
$2p_{3/2}$
     & SESE & $-0.057361\,(m/m_r)$ & $ $0$   $ &  $0$    &   $1/9$      & $-0.047780$ \\
     & SEVP & $ 0.002615\,(m/m_r)$ & $ $0$   $ &  $0$    &  $0$       & $-2/45$ \\
     & VPVP & $0        $& $ $0$    $ &  $0$    &  $0$       & $0$\\[5pt]
\end{tabular}
\end{ruledtabular}
\end{table*}

\subsubsection{Two-loop QED correction}

The two-loop QED correction is parameterized as
\begin{align}\label{qed2:1}
E_{\rm QED2} = mc^2\,\left( \frac{m_r}{m}\right)^3\, \left(\frac{\alpha}{\pi}\right)^2\, \frac{(\Za)^4}{n^3}\, F_{\rm QED2}(\Za)\,,
\end{align}
where $F_{\rm QED2}(\Za)$ is a slowly-varying dimensionless function of $\Za$. The $\Za$-expansion
of the function $F_{\rm QED2}$ is given by \cite{sapirstein:90:kin}
\begin{align} \label{eq:twoloop:za}
F_{\rm QED2}(\Za) =& \  B_{40} + (\Za)\,B_{50} + (\Za)^2\bigl[ B_{63}\, L^3
  \nonumber \\ &
+ B_{62}\, L^2 + B_{61}\, L + G_{\rm QED2}(\Za)\bigr]\,,
\end{align}
where $G_{\rm QED2}(\Za) = B_{60} + \ldots$ is the remainder that contains all higher-order
expansion terms in $\Za$.

The two-loop QED correction is represented by the set of Feynman diagrams shown in
Fig.~\ref{fig:2order}. In the present compilation we divide the two-loop QED correction into three
parts: the two-loop self-energy correction $E_{\rm SESE}$ (diagrams in Fig.~\ref{fig:2order} a-c),
the two-loop vacuum-polarization correction $E_{\rm VPVP}$ (diagrams in Fig.~\ref{fig:2order} g-i),
and the mixed self-energy and vacuum-polarization correction $E_{\rm SEVP}$ (diagrams in
Fig.~\ref{fig:2order} d, e, f, and k),
\begin{align} \label{qed2:2}
E_{\rm QED2} = E_{\rm SESE} + E_{\rm SEVP} + E_{\rm VPVP}\,.
\end{align}
Coefficients of the $\Za$ expansion of the individual two-loop corrections for the states under
consideration are summarized in Table~\ref{tab:twoloop}, for details see
Refs.~\cite{pachucki:01:pra,pachucki:03:prl,jentschura:05:sese} and references to the earlier works
therein.

The {\em two-loop self-energy correction} was calculated to all orders in $\Za$ in
Refs.~\cite{yerokhin:03:prl,yerokhin:03:epjd,yerokhin:05:sese,yerokhin:05:jetp,yerokhin:09:sese,yerokhin:10:sese}
for the $1s$ state and in Refs.~\cite{yerokhin:06:prl,yerokhin:07:cjp} for the $n=2$ states.
Numerical all-order results are available for $Z \ge 10$ for the $1s$ state and for $Z \ge 60$ for
the $n = 2$ states. An extrapolation of all-order numerical values towards $Z = 1$ was presented in
Ref.~\cite{yerokhin:09:sese} for the $1s$ state, in marginal agreement with the $\Za$ expansion
results \cite{pachucki:03:prl}.

Numerical values of the SESE correction in the present compilation were obtained as follows. For
the $1s$ state and $Z \ge 10$, we interpolate the numerical values of the higher-order remainder
$G_{\rm SESE}$ extracted from all-order results of
Refs.~\cite{yerokhin:05:sese,yerokhin:09:sese,yerokhin:10:sese}. For the $1s$ state and $Z < 10$,
we extrapolate the higher-order remainder $G_{\rm SESE}$ as described in
Ref.~\cite{yerokhin:09:sese}. In the case of hydrogen, the value of the higher-order remainder and
its uncertainty are consistent with the results of Ref.~\cite{yerokhin:09:sese}. For the $2s$
state, we divide the higher-order remainder $G_{\rm SESE}(2s)$ into two parts,
$$G_{\rm SESE}(2s) = G_{\rm SESE}(1s) + \bigl[ G_{\rm SESE}(2s) - G_{\rm SESE}(1s) \bigr]\,.$$  The
difference $G_{\rm SESE}(2s) - G_{\rm SESE}(1s)$ is obtained by interpolating the numerical results
\cite{yerokhin:06:prl} combined with the $Z = 0$ limiting value $B_{60}(2s)-B_{60}(1s) = 14.1(4)$
\cite{jentschura:05:sese}. For the $2p_j$ states and $Z \ge 60$, the compilation results are
obtained by interpolating the numerical data for the higher-order remainder \cite{yerokhin:06:prl}.
For the $2p_j$ states and $Z < 60$, we interpolate the numerical results combined with the $Z = 0$
limiting value $B_{60}(2p) = -2.2(3)$ \cite{pachucki:03:prl}.

Nonperturbative (in $\Za$) calculations of the {\em two-loop vacuum-polarization correction} and
the {\em mixed self-energy and vacuum-polarization correction} were performed in
Refs.~\cite{persson:96:pra,mallampalli:96,beier:97:jpb,plunien:98:epj} and later extended in
Ref.~\cite{yerokhin:08:twoloop}. These calculations are not fully completed yet since several
diagrams (specifically, those  in Fig.~\ref{fig:2order} i and k) are computed within the free-loop
approximation (see Ref.~\cite{yerokhin:08:twoloop} for details). The uncertainty due to this
approximation induces the largest theoretical error in the Lamb shift (which competes with the
error induced by the uncertainty of the nuclear charge radii).

In the present compilation, the numerical results for the SEVP and VPVP corrections are taken from
Ref.~\cite{yerokhin:08:twoloop}. In order to reduce the uncertainty of interpolation, we subtracted
from the all-order results of Ref.~\cite{yerokhin:08:twoloop} the known terms of the
$\Za$-expansion (\ref{eq:twoloop:za})  and interpolated the higher-order remainders, $G_{\rm
SEVP}(\Za)$ and $G_{\rm VPVP}(\Za)$.

The uncertainty of the SEVP correction comes from the uncalculated effects of relative order
$(\Za)^{6+}$ beyond the free-loop approximation. In order to estimate this uncertainty for the $s$
states, we take the contribution of order $(\Za)^{6+}$ in the free-loop approximation, multiply it
by $1.69$, which is the ratio of the $(\Za)^5$ terms beyond the free-loop approximation and within
this approximation, and by the conservative factor of $1.5$. For the $p$ states, the $(\Za)^5$
contribution is zero, so we estimate the uncertainty as the absolute value of the $(\Za)^{6+}$
contribution in the free-loop approximation multiplied by the conservative factor of 1.5.

For the VPVP correction, we estimate the uncertainty by the absolute value of the (free-loop
approximation) contribution of the diagrams in Fig.~\ref{fig:2order}(h and i) multiplied by a
factor of $(\Za)^2$. We note that the diagram Fig.~\ref{fig:2order}(h) was calculated beyond the
free-loop approximation for uranium and lead in Ref.~\cite{plunien:98:epj}. The additional
correction found in that work is rather small; its value is consistent with our definition of
uncertainty.

\subsubsection{Higher-order QED corrections}

We now collect the remaining higher-order QED corrections. Their contribution to the Lamb shift is
very small and relevant for the very low-$Z$ ions only. Because of this, we include these
corrections in our tabulation for $Z\le 10$ only.

The three-loop QED correction is known to its leading order in the $\Za$ expansion,
\begin{align} \label{qed3:1}
E_{\rm QED3} = mc^2\,\left(\frac{\alpha}{\pi}\right)^3\, \frac{(\Za)^4}{n^3}\, C_{40}\,,
\end{align}
where the coefficient $C_{40}$ has the following numerical values for the states of the present
interest (see Ref.~\cite{mohr:12:codata} and references in there),
\begin{align}
C_{40}(ns) &\ = 0.417504\,,\\
C_{40}(np_{1/2}) &\ = -0.393747\,,\\
C_{40}(np_{3/2}) &\ = 0.196874\,.
\end{align}
Following Ref.~\cite{mohr:12:codata}, we assign the uncertainty of $30\Za\,\delta_{l,0}$ in units
of $C_{40}$ to this correction.

The vacuum-polarization induced by the $\mu^+\mu^-$ virtual pairs is given by
\cite{eides:95:pra,karshenboim:95}
\begin{align}\label{qed3:2}
E_{\mu \rm VP} = mc^2\, \left(\frac{m}{m_{\mu}}\right)^2 \frac{\alpha}{\pi}\,\frac{(\Za)^4}{n^3}\,
\left( -\frac{4}{15}\,\delta_{l,0}\right)\,.
\end{align}

The hadronic vacuum-polarization correction is of the same order as the muonic vacuum polarization
\cite{friar:99},
\begin{align}\label{qed3:3}
E_{ \rm hadVP} = 0.671\,(15)\ E_{\mu \rm VP}\,.
\end{align}

\subsubsection{Nuclear size correction to the Dirac energy}

The finite nuclear size (FNS) correction to the Dirac energy is parameterized in terms of the
dimensionless function ${G}_{\rm FNS}$ as \cite{shabaev:93:fns}
\begin{align} \label{eq:fns}
E_{\rm FNS} = mc^2\, \left(\frac{m_r}{m}\right)^3
  \left( \frac{2\,\Za\,R_{\rm Sph}}{n\,\lambdabar_C} \right)^{2\gamma}\,
        \frac{(\Za)^K}{n}\,
    {G}_{\rm FNS}(\Za,R)\,,
\end{align}
where $R_{\rm Sph} = \sqrt{5/3}\,R$ is the radius of the sphere with the root-mean-square (rms)
radius $R$, $\gamma = \sqrt{1-(\Za)^2}$, and $\lambdabar_C$ is the Compton wavelength divided by $
2\pi$. The power of the leading $Z$ dependence is $K = 2$ for $ns$ states and $K = 4$ for $np$
states.

A numerical calculation of ${G}_{\rm FNS}(\Za,R)$ is quite straightforward for most ions. In the
present tabulations we use results of our own calculations. They were performed by solving the
Dirac equation with two independent methods, the one based on the Dual Kinetic Balance approach
\cite{shabaev:04:DKB} and the other based on the RADIAL package \cite{salvat:95:cpc}. Comparison of
the results obtained by these two methods was used as a check of numerical accuracy. For low-$Z$
ions, the FNS contribution becomes very small, and the calculations are complicated by large
numerical cancellations. In order to overcome this problem, we implemented the Dual Kinetic Balance
method in the quadruple (32-digit) arithmetics, which allowed us to perform accurate numerical
determination of the FNS corrections for all ions including hydrogen. Numerical values obtained in
this way were checked against the results obtained by analytical formulas from
Ref.~\cite{shabaev:93:fns}.

In our calculations, we used three models of the nuclear charge distribution. The two-parameter
Fermi model is given by
\begin{align}\label{fermi}
\rho_{\rm Fer}(r) = \frac{N_{0}}{1+\exp[(r-r_{0})/a]}\,,
\end{align}
where $r_0$ and $a$ are the parameters of the Fermi distribution, and $N_0$ is the normalization
factor. The parameter $a$ was fixed by the standard choice of $a = 2.3/(4\ln3)\approx 0.52$~fm. For
a given value of the rms radius $R$, the parameter $r_0$ was determined by the formula
\begin{align} \label{2}
r_0^2 = \frac{5}{3} R^2-\frac{7}{3} a^2 \pi^2\,.
\end{align}
Though this formula is approximate, for the purposes of the present work it was sufficient to
determine the parameter $r_0$ just from this formula.

The homogeneously charged sphere distribution of the nuclear charge is given by
\begin{align} \label{1}
\rho_{\rm Sph}(r) = \frac{3}{4\pi R^3_{\rm Sph}}\,\theta(R_{\rm Sph}-r)\,,
\end{align}
where $\theta$ is the Heaviside step function. The Gauss distribution of the nuclear charge reads
\begin{align} \label{1a}
\rho_{\rm Gaus}(r) = \left( \frac{3}{2\pi R^2 }\right)^{3/2}\,\exp\left(-\frac{3\, r^2}{2R^2}\right)\,.
\end{align}

For all ions with $Z\ge 10$ except uranium, we performed calculations of the FNS correction with
the Fermi model of the nuclear charge distribution, and estimated the model dependence by taking
the difference of these results and the ones obtained with the homogeneously charged sphere model,
as suggested in Ref.~\cite{franosch:91}. For uranium, we used the results from
Ref.~\cite{kozhedub:08}, which were obtained with a more realistic nuclear model. For $Z < 10$, the
Fermi model  of the nuclear charge distribution is no longer adequate, so we used the Gauss model
instead and estimated the model dependence by taking the difference of those results and the ones
obtained with the homogeneously charged sphere model. For hydrogen, our numerical values are
consistent with the results of Ref.~\cite{mohr:12:codata} based on the $\Za$ expansion approach.

For some ions, parameters of the nuclear charge distribution are known with a high accuracy, and so
the inclusion of the effect of the nuclear shape (deformation) makes sense. For most nuclei the
nuclear deformation correction is  small; we estimate it to be within the model dependence error
evaluated as described above. The nuclear deformation effect is explicitly taken into account only
for $^{238}$U, as calculated in Ref.~\cite{kozhedub:08}. For this ion, we include the nuclear
deformation correction of $-0.14$~eV, $-0.026$~eV, and $-0.003$~eV for the $1s$, $2s$, and
$2p_{1/2}$ states, correspondingly. The uncertainty of the FNS correction for $^{238}$U was also
taken from Ref.~\cite{kozhedub:08}.

\subsubsection{Nuclear size corrections to QED effects}

The finite nuclear size effect on the one-loop QED corrections is parameterized as
\cite{milstein:04,yerokhin:11:fns}
\begin{align} \label{eq:sefns}
E_{\rm QED1, fns} = E_{\rm FNS}(n\nicefrac12 l)\,\frac{\alpha}{\pi}\,
    {G}_{\rm QED1, fns}(\Za,R)\,,
\end{align}
where $E_{\rm FNS}(n j l)$ is the FNS correction (\ref{eq:fns}) for the state with the quantum
numbers $n$, $j$, and $l$ and ${G}_{\rm QED1, fns}$ is a slowly varying function of $\Za$ and $R$.
Note that the value of $j$ in $E_{\rm FNS}$ in the right-hand-side of Eq.~(\ref{eq:sefns}) is fixed
by $j = 1/2$. This means that for the $np_{3/2}$ reference state, Eq.~(\ref{eq:sefns}) has the FNS
correction for the $np_{1/2}$ state, $E_{\rm FNS}(np_{1/2})$, as suggested in
Ref.~\cite{milstein:04}.

Three one-loop QED corrections given by Eqs.~(\ref{se:1}), (\ref{ueh:1}), and (\ref{wk:1}) induce,
correspondingly, three finite nuclear size contributions,
\begin{align}\label{QED1fns}
E_{\rm QED1, fns} = E_{\rm SE, fns} + E_{\rm Ueh, fns} + E_{\rm WK, fns}\,.
\end{align}
First calculations of the FNS effect on the self-energy and vacuum-polarization were performed,
correspondingly, in Refs.~\cite{mohr:93:prl} and \cite{soff:88:vp} for the hollow-shell model of
the nuclear charge distribution. Calculations for more realistic nuclear models were later
performed in other studies, notably, in Refs.~\cite{beier:98:pra,yerokhin:11:fns}.

In the present compilation we obtain the FNS correction to the electron self-energy by
interpolating the results for ${G}_{\rm SE, fns}$  for the Fermi nuclear charge distribution from
Ref.~\cite{yerokhin:11:fns}. In several cases we performed additional calculations by the same
method in order to reduce interpolation errors. Estimating the uncertainty due to the model
dependence of this correction, we accounted only for the leading error coming from the prefactor
$\Delta E_{\rm FNS}(n\nicefrac12 l)$ in Eq.~(\ref{eq:sefns}).

The finite nuclear size correction to the one-loop Uehling contribution is relatively simple and
can be calculated with a very high accuracy \cite{yerokhin:11:fns}. In the present compilation we
list the results of our own calculations of this correction for the Fermi nuclear charge
distribution.

The finite nuclear size correction to the one-loop Wichmann-Kroll contribution was obtained by
interpolating the results for the function ${G}_{\rm WK, fns}$  for the Fermi nuclear charge
distribution from Ref.~\cite{yerokhin:11:fns}.

\subsubsection{Nuclear mass (recoil) corrections}
\label{sec:recoil}

The leading nuclear mass (recoil) contribution is separated out from the Lamb shift as the $E_{\rm
M}$ correction in Eq.~(\ref{eq1a}). Apart from $E_{\rm M}$, we have already accounted for the
reduced-mass dependence in the self-energy correction [Eq.~(\ref{se:1})], vacuum-polarization
correction [Eqs.~(\ref{ueh:1}) and (\ref{wk:1})], the two-loop QED correction [Eq.~(\ref{qed2:1})],
and the finite nuclear size correction [Eq.~(\ref{eq:fns})]. In this section we summarize the
remaining recoil contributions.

The {\em first-order (in $m/M$) recoil correction} is represented as
\begin{align}\label{rec:1}
E_{\rm REC} = mc^2\,\frac{m}{M}\, \frac{(\Za)^5}{\pi\, n^3}\,P(\Za)\,,
\end{align}
where
\begin{align}
P(\Za) = &\ \ln (\Za)^{-2}\,D_{51} + D_{50}
  \nonumber \\ &
 + (\Za)\, D_{60}+ (\Za)^2\,G_{\rm REC}(\Za)\,,
\end{align}
and $G_{\rm REC}(\Za)$ is the higher-order remainder containing all higher orders in $\Za$. The
coefficients of the $\Za$ expansion are (see Ref.~\cite{mohr:12:codata} and references therein)
\begin{align}
D_{51}  = &\ \left( \frac{m_r}{m}\right)^3\,\frac13\,\delta_{l,0}\,,\\
D_{50}  = &\ \left( \frac{m_r}{m}\right)^3\,\biggl[ -\frac83\,\ln k_0(n,l) + d_{50}\biggr] \,,\\
D_{60}  = &\ \left( 4\ln 2-\frac{7}{2}\right)\pi\,\delta_{l,0}
  \nonumber \\ &
+ \left[ 3- \frac{l(l+1)}{n^2}\right]\, \frac{2\pi(1-\delta_{l,0})}{(4l^2-1)(2l+3)}\,,
\end{align}
where the values of $d_{50}(n,l)$ for the states of interest are
\begin{align}
d_{50}(1s) &\ = \frac{14}{3}\ln 2 + \frac{62}{9}\,,\\
d_{50}(2s) &\ = \frac{187}{18}\,,\\
d_{50}(2p) &\ = -\frac{7}{18}\,.
\end{align}

The higher-order remainder $G_{\rm REC}(\Za)$ is extracted from the numerical all-order
calculations. The results listed in the present tabulations are obtained for $Z > 5$ by
interpolating the higher-order remainder inferred from the all-order results taken from
Refs.~\cite{artemyev:95:pra,shabaev:98:recground} for the $1s$, $2s$ and $2p_{1/2}$ states and from
Ref.~\cite{artemyev:95:jpb} for the $2p_{3/2}$ state. For $Z <= 5$, there is a small inconsistency
in the numerical all-order results \cite{artemyev:95:pra,shabaev:98:recground,shabaev:98:jpb} and
the analytical result for the $D_{72}$ coefficient \cite{pachucki:99:prab,melnikov:99}. Because of
this, the numerical values listed in the tabulation for $Z <= 5$ were obtained by the polynomial
extrapolation of the all-order results for $G_{\rm REC}(\Za)$ with $Z > 5$. A conservative
uncertainty was assigned to these values, which preserves consistency with the result obtained for
hydrogen in Ref.~\cite{shabaev:98:jpb}.

The finite nuclear size correction to the recoil effect $E_{\rm REC,FNS}$ was calculated within
some approximation in Refs.~\cite{shabaev:98:recground,shabaev:98:ps,aleksandrov:15}. In the
present compilation, we include this correction for $Z\ge 10$. For the $1s$ and $2s$ states, we
interpolate the numerical results from Refs.~\cite{shabaev:98:recground,shabaev:98:ps}, after
removing the reduced-mass contribution, which is already taken into account in Eq.~(\ref{eq:fns}).
The uncertainty ascribed to this contribution comprises the numerical and the interpolation errors
as well as the error of the approximation. The latter error was evaluated for the $s$ states by
taking the relative deviation of the exact results for the low-order recoil part
\cite{aleksandrov:15} from the results of the approximate treatment
\cite{shabaev:98:recground,shabaev:98:ps}. For the $2p_{1/2}$ state, we include only the low-order
part of the correction, as obtained in Ref.~\cite{aleksandrov:15}. The uncertainty was estimated by
multiplying $E_{\rm REC}(2p_{1/2})$ with the ratio $E_{\rm REC,FNS}(2s)/E_{\rm REC}(2s)$. For the
$2p_{3/2}$ state, we assume the correction to be zero and estimate the uncertainty analogously to
that for the $2p_{1/2}$ state.

The {\em radiative recoil correction} is given by (see Ref.~\cite{mohr:12:codata} and references
therein)
\begin{align}\label{rec:2}
E_{\rm RREC} = &\ mc^2\,\frac{m}{M}\, \frac{\alpha(\Za)^5}{\pi^2 n^3}\,\delta_{l,0}
\biggl[ 6\,\zeta(3)-2\pi^2\ln 2
  \nonumber \\ &
+ \frac{35\pi^2}{36} - \frac{448}{27} + \frac{2}{3}\,\pi(\Za)\ln^2(\Za)^{-2} \biggr]\,,
\end{align}
where $\zeta$ is the Riemann zeta function.  Following Ref.~\cite{mohr:12:codata}, we ascribe to
this correction an uncertainty of $10(\Za)\ln(\Za)^{-2}$ relative to the square brackets in the
above equation. Because of the smallness of the radiative recoil, it is included into the
compilation only for $Z\le 10$.

Relativistic recoil correction of order $(\Za)^4 (m/M)^2$ depends on the nuclear spin and is known
for nuclei with spin $I = 0$, $1/2$, and $1$ \cite{pachucki:95:jpb},
\begin{align}\label{rec:3}
E^{(4)}_{\rm REC,2} = mc^2\,\left(\frac{m}{M}\right)^2\, \frac{(\Za)^4}{2 n^3}\,
 \left[ \frac{3}{2n} - \frac2{2l+ 1} + \delta_{l,0}\,\delta_{I,1/2}\right]\,.
\end{align}
The last term in the square brackets in the above expression is nonzero only for states with $l=0$
and the nuclear spin $I = 1/2$. The correction $E^{(4)}_{\rm REC,2}$ is very small and relevant
only for hydrogen and He$^+$.

The relativistic recoil correction of order $(\Za)^5 (m/M)^2$ is extremely small and relevant for
hydrogen only. It is given by \cite{sapirstein:90:kin}
\begin{align}\label{rec:4}
E^{(5)}_{\rm REC,2} = mc^2\,\left(\frac{m}{M}\right)^2\, \frac{(\Za)^5}{\pi n^3}\,2\,\delta_{l,0}
\left[ \frac{m}{M}\ln \frac{M}{m}-1\right]\,.
\end{align}

\subsubsection{Nuclear polarization and self-energy}

The nuclear polarization correction has been calculated in the literature for selected ions only.
We take the numerical results for $Z = 1$ from Ref.~\cite{khriplovitch:00}, for $Z = 2$ from
Ref.~\cite{pachucki:07:heliumnp}, for $Z = 82$ and $92$ from Ref.~\cite{nefiodov:96}, and for $Z =
90$, $94$, $96$, and $98$ from Refs.~\cite{plunien:89,plunien:91,plunien:95}.

For $Z = 36$, $44$, $48$, $60$, $64$, $66$, $70$, $78$, and $80$, the nuclear polarization
correction to the Lamb shift can be deduced from the calculations of this correction for the
bound-electron $g$ factor \cite{nefiodov:02:prl}. We obtain it by the following approximate
relation
\begin{align}\label{np}
E_{\rm NP} \approx mc^2\,\frac{j(j+1)}{3\kappa^2}\,\delta g_{\rm NP}\,,
\end{align}
where $E_{\rm NP}$ is the nuclear polarization correction to the Lamb shift and $\delta g_{\rm NP}$
is the nuclear polarization correction to the $g$ factor. Relation (\ref{np}) can be easily
obtained by using the generalized virial relations \cite{shabaev:03:PSAS} for any
spherically-symmetric perturbation potential with a sufficiently short range. We ascribe a
numerical uncertainty of 50\% to all above mentioned results for the nuclear polarization.

For the ions for which no calculations of the nuclear polarization correction were performed, we
use the following conservative estimate
\begin{align}
E_{\rm NP} \approx - \frac1{1000}\,E_{\rm FNS}\,,
\end{align}
where $E_{\rm FNS}$ is the finite nuclear-size correction. We ascribe the uncertainty of 100\% to
this estimation.

The nuclear self-energy correction is a tiny contribution relevant only for hydrogen
\cite{pachucki:95:pra},
\begin{align}\label{nse}
E_{\rm NSE} = &\ mc^2\,\left(\frac{m}{M}\right)^2\, \frac{4Z(\Za)^5}{3\pi n^3}\,
  \nonumber \\ &
\times \biggl[ \ln\left(\frac{M}{m(\Za)^2}\right)\delta_{l,0}
- \ln k_0(n,l)
\biggr]\,.
\end{align}
Following Ref.~\cite{mohr:12:codata}, we estimate the uncertainty of this correction as 0.5 in the
square brackets in the above formula.

\section{Tabulations}

The following values of fundamental constants were used in the present compilation
\cite{mohr:12:codata}:
\begin{align}
\alpha^{-1} &\ = 137.035\,999\,074\,(44)\,,\\
m &\ = 5.485\,799\,0946\,(22)\,\times 10^{-4}\ \mbox{\rm a.m.u.}
\end{align}

The Table~\ref{tab:lamb} lists the individual contributions to the Lamb shift of H-like ions. The
contributions are presented in terms of the function $F(\Za)$ defined as
\begin{align}
E = mc^2\,\frac{\alpha}{\pi}\, \frac{(\Za)^4}{n^3}\, F(\Za)\,.
\end{align}

For each element, an isotope is selected (in most cases, the most abundant one), for which the
nuclear parameters were taken. The nuclear masses were taken from the Ame2012 compilation
\cite{wang:12}. In order to convert the atomic excess mass $M_{\rm A, ex}$ tabulated in that work
to the mass of the nucleus, the following relation was used (in atomic mass units)
\begin{align}
M = M_{\rm A, ex} + A - m\,Z + B_e(Z)\,,
\end{align}
where $A$ is the atomic mass number and $B_e(Z)$ is the total binding energy of all electrons in
the atom. The binding energy $B_e(Z)$ was estimated by the approximation formula from
Ref.~\cite{lunney:03}  (in eV),
\begin{align}
B_e(Z) = 14.4381\, Z^{2.39} + 1.55468\times 10^{-6}\, Z^{5.35}\,.
\end{align}
After checking this equation against the Dirac-Fock calculations of the total binding energy in
Ref.~\cite{rodrigues:04}, we assigned the uncertainty of 2\% to this formula.

Experimental values of the root-mean-square (rms) nuclear charge radii used in the compilation were
taken from Ref.~\cite{angeli:13}. In the rare cases when no experimental data were available, we
used the standard approximation formula \cite{johnson:85} (in fermi)
\begin{align}
R = 0.836\,A^{1/3} + 0.570\,,
\end{align}
and ascribed the uncertainty of 10\% to this value of $R$.

The uncertainty of the finite nuclear size corrections listed in Table~\ref{tab:lamb} is separated
into two parts. The first part represents the estimated error due to the nuclear model dependence
and the numerical error (if any). The second part is the uncertainty due to the error of the
nuclear rms charge radius $R$, which is easily obtained as
\begin{align}
\delta E_{\rm FNS} = E_{\rm FNS}\, 2\gamma \frac{\delta R}{R} \,,
\end{align}
where $\delta R$ is the uncertainty of the radius $R$, $\gamma = \sqrt{1-(\Za)^2}$.

We now describe the individual corrections listed in Table~\ref{tab:lamb} in detail.

\begin{tabular}{ll}
\\
SE(pnt)& \begin{minipage}[t]{0.75\columnwidth}
    One-loop self-energy correction for the point nucleus in the
    nonrecoil limit. It is given by Eq.~(\ref{se:1}) with the removed reduced-mass dependence, $m_r/m \to
    1$. The specified uncertainty (if any) represents the numerical error of the
    calculation.\\
    \end{minipage}\\
Ueh(pnt)& \begin{minipage}[t]{0.75\columnwidth}
    One-loop Uehling vacuum-polarization correction for the point
    nucleus in the nonrecoil limit. It is given by Eq.~(\ref{ueh:1}) with the removed reduced-mass dependence, $m_r/m \to 1$.\\
    \end{minipage}
    \\
WK(pnt)& \begin{minipage}[t]{0.75\columnwidth}
    One-loop Wichmann-Kroll vacuum-polarization correction for the
    point nucleus in the nonrecoil limit. It is given by Eq.~(\ref{wk:1}) with the removed reduced-mass dependence, $m_r/m \to 1$.\\
    \end{minipage}
    \\
FNS & \begin{minipage}[t]{0.75\columnwidth}
    The finite nuclear size correction to the Dirac energy, in the nonrecoil limit.
    It is given by Eq.~(\ref{eq:fns}) with the removed reduced-mass dependence, $m_r/m \to 1$. The first
    uncertainty represents the nuclear-model dependence of this correction; the second uncertainty is due to
    the error of the nuclear charge radius. \\
    \end{minipage}
    \\
\end{tabular}
\begin{tabular}{ll}
\\
SE(fns) & \begin{minipage}[t]{0.75\columnwidth}
    Finite-nuclear size correction to the electron self-energy, given by the first term in Eq.~(\ref{QED1fns}).
    The first uncertainty comprises
    the numerical error and the estimation of the model dependence; the second uncertainty is due to
    the error of the nuclear charge radius.\\
    \end{minipage}
    \\
Ueh(fns) & \begin{minipage}[t]{0.75\columnwidth}
    Finite-nuclear size correction to the Uehling vacuum-polarization, given by the second term in Eq.~(\ref{QED1fns}). The first
    uncertainty is the estimation of the model dependence; the second uncertainty is due to
    the error of the nuclear charge radius.\\
    \end{minipage}
    \\
WK(fns) & \begin{minipage}[t]{0.75\columnwidth}
    Finite-nuclear size correction to the Wichmann-Kroll
    vacuum-polarization, given by the third term in Eq.~(\ref{QED1fns}). The first
    uncertainty includes the numerical error and the estimation of the model dependence; the second uncertainty is due to
    the error of the nuclear charge radius.\\
    \end{minipage}
    \\
SESE & \begin{minipage}[t]{0.75\columnwidth}
    The two-loop self-energy correction, given by the first term
    in Eq.~(\ref{qed2:2}). The quoted uncertainty includes the interpolation (extrapolation) error
    and the numerical error.\\
    \end{minipage}
    \\
SEVP & \begin{minipage}[t]{0.75\columnwidth}
    The two-loop mixed self-energy-vacuum-polarization correction, given by the second term in Eq.~(\ref{qed2:2}).
    The quoted uncertainty is the estimation of
    the uncalculated higher-order effects.\\
    \end{minipage}
    \\
VPVP & \begin{minipage}[t]{0.75\columnwidth}
    The two-loop vacuum-polarization correction, given by the
    third term in Eq.~(\ref{qed2:2}). The quoted uncertainty is the estimation of the
    uncalculated higher-order effects.\\
    \end{minipage}
    \\
RRM & \begin{minipage}[t]{0.75\columnwidth}
    Relativistic reduced-mass recoil correction. It is given by the sum of the
    reduced-mass dependence of the one-loop self-energy correction [Eq.~(\ref{se:1})], the
    one-loop vacuum-polarization correction [Eqs.~(\ref{ueh:1}) and (\ref{wk:1})], the two-loop QED correction
    [Eq.~(\ref{qed2:1})], and the finite nuclear size [Eq.~(\ref{eq:fns})].\\
    \end{minipage}
    \\
REC & \begin{minipage}[t]{0.75\columnwidth}
    First-order recoil correction, Eq.~(\ref{rec:1}). The quoted uncertainty includes the interpolation (extrapolation) error
    and the numerical error.\\
    \end{minipage}
    \\
NUCL & \begin{minipage}[t]{0.75\columnwidth}
    Nuclear polarization and nuclear self-energy corrections,
    Eqs.~(\ref{np})-(\ref{nse}). The quoted uncertainty is the estimation of the approximation error.\\
    \end{minipage}
    \\
REC(fns) & \begin{minipage}[t]{0.75\columnwidth}
    Finite nuclear size correction to the recoil effect. Listed for $Z\ge 10$ only. The quoted uncertainty includes
    the numerical and interpolation errors as well as the estimation of the approximation error.\\
    \end{minipage}
    \\
\end{tabular}
\begin{tabular}{ll}
\\
REC(ho) & \begin{minipage}[t]{0.75\columnwidth}
    Higher-order recoil correction, given by the sum of Eqs.~(\ref{rec:2}), (\ref{rec:3}),
    and (\ref{rec:4}). Listed for $Z\le 10$ only. The quoted uncertainty is the estimation of the uncalculated higher-order effects.\\
    \end{minipage}
    \\
QED(ho) & \begin{minipage}[t]{0.75\columnwidth}
    Higher-order QED correction, given by the sum of Eqs.~(\ref{qed3:1}), (\ref{qed3:2}),
    and (\ref{qed3:3}). Listed for $Z\le 10$ only. The quoted uncertainty is the estimation of the uncalculated higher-order effects.\\
    \end{minipage}
    \\
Total & \begin{minipage}[t]{0.75\columnwidth}
    The total values of the Lamb shift, given by the sum of all contributions. The second uncertainty is due to
    the error of the nuclear charge radius; the first uncertainty is the sum of all other errors of individual contributions, added quadratically.
\\
    \end{minipage}
    \\
\end{tabular}

In Table~\ref{tab:energy} we present the binding energies of the $1s$ state and the energy level
separation intervals $2s-1s$, $2p_{1/2}-2s$, and $2p_{3/2}-2p_{1/2}$ for hydrogenlike atoms with $1
\le Z \le 110$. All energies are presented as a sum of three parts that correspond to the three
terms in the right-hand-side of Eq.~(\ref{eq1a}).

It might be instructive to compare the individual contributions to the Lamb shift in the present
compilation with the corresponding results from the compilation by Johnson and Soff
\cite{johnson:85} as tabulated 30 years ago. Such comparison for three selected ions with $Z = 5$,
$50$, and $92$ is shown in Table~\ref{tab:js}. For the comparison, we re-grouped the contributions
from Table~\ref{tab:lamb} in the same way as in Ref.~\cite{johnson:85}.

We observe that the main change in the theoretical values for the total Lamb shift in medium- and
high-$Z$ region is due to the two-loop QED corrections, which were unknown at the time of the
previous compilation. After the calculations \cite{yerokhin:01:sese,yerokhin:03:prl} have been
carried out, both the sign and the magnitude of the two-loop QED contribution turned out to be
different from what had been estimated. Another important difference comes from the FNS correction,
which is due to changes in the experimental values of the nuclear charge radii. Apart from that, we
observe good agreement with the results from Ref.~\cite{johnson:85} for most of the contributions.
The total theoretical accuracy of the Lamb shift is considerably improved (typically, by a factor
of 5) as compared to the previous compilation.

The results for hydrogen in Tables~\ref{tab:lamb} and \ref{tab:energy} are given mainly for
completeness, since the hydrogen theory is critically reviewed and regularly updated in the CODATA
adjustment of fundamental constants \cite{mohr:12:codata}. The theoretical values of the hydrogen
transition energies listed in Table~\ref{tab:energy} are consistent with the CODATA 2010 adjustment
\cite{mohr:12:codata} and in agreement with the (much more accurate) Garching experimental result
for the $2s-1s$  transition in hydrogen \cite{fischer:04}. The comparison with hydrogen experiments
cannot be used as a test of theoretical calculations, since hydrogen spectroscopy is used for the
determination of the Rydberg constant and the proton charge radius.

In Table~\ref{tab:comparison} we compare the theoretical predictions for the Lamb shift and
transition energies in hydrogenlike ions with the available experimental results.

The Lamb shift for the $1s$ state is much larger than for the excited states but difficult to
access in an experiment. It is usually determined experimentally by studying $np-1s$ transitions.
Since the Lamb shift forms only a small part of these transition energies, very precise
measurements in the X-ray regime are required. The highest fractional accuracy so far was achieved
in the experiment on U$^{92+}$ \cite{gumberidze:05:prl}, which has reached the sensitivity of 1\%
to the $1s$ Lamb shift. Accurate measurements were reported also for other ions, notably, for
Ar$^{17+}$ \cite{kubicek:14} and for Ni$^{27+}$ \cite{beyer:91}, with the fractional errors of
1.2\% and 2.0\% of the Lamb shift, correspondingly.

The $2s$ Lamb shift is approximately eight times smaller than the $1s$ Lamb shift, but it is easier
to access experimentally since the $2s$ and the $2p$ levels are close in energy. The $2p_{1/2}-2s$
and $2p_{3/2}-2s$ transitions are accessible to laser spectroscopy over a wide region of $Z$ using
far-infrared to ultraviolet lasers. A serious problem, however, is the natural width of the
transitions that grows with $Z$ as $Z^4$. Because of this, accurate experimental results for
$2p-2s$ transitions are available mostly for low- and medium-$Z$ ions.

One of the most accurate experiments of this kind is the measurement of the classical $2p_{1/2}-2s$
Lamb shift in He$^+$ \cite{wijngaarden:00}, which reached the sensitivity to QED effects on the
level of $12$ parts per million. Unfortunately, this measurement disagrees with theory by about two
standard deviations. Accurate measurements of the $2s$ Lamb shift have been also reported for
$^{14}$N$^{6+}$ \cite{myers:01}, $^{31}$P$^{14+}$ \cite{pross:93}, and $^{32}$S$^{15+}$
\cite{georgiadis:86}, with a precision of $0.1$-$0.3\%$ of the Lamb shift.

\section*{Conclusion}


In this paper we tabulated theoretical energies and energy-level separations of $n = 1$ and $n = 2$
states of the hydrogenlike atoms with the nuclear charges $1 \le Z \le 110$. A detailed breakdown
of individual QED corrections to the Lamb shift of the $1s$, $2s$, $2p_{1/2}$, and $2p_{3/2}$
states is provided for each atom.

The main sources of uncertainty of the theoretical energy levels in hydrogenic systems are
presently the two-loop QED corrections and the lack of detailed knowledge of the nuclear charge
distributions. Both these problems can be addressed in future in theoretical and experimental
studies. Principal limitations of the theoretical description are associated with the dynamical
nuclear effects, such as the nuclear polarization. Accurate calculations of such effects would
require detailed understanding of the nuclear structure, which is presently lacking. So far,
however, the uncertainty due to nuclear polarization is not the dominant source of the theoretical
error for most of the hydrogenlike atoms. It is therefore likely that the precision of the
theoretical description of the hydrogenlike systems will be improved further in the future.

\section*{Acknowledgement}

We are grateful to A.~A.~Nekipelov for providing us the code for the computation of the
Wichmann-Kroll potential. V.A.Y. acknowledges support by the Russian Federation program for
organizing and carrying out scientific investigations. The work of V.M.S. was supported by RFBR
(grant No. 13-02-00630) and by SPbSU (grants No. 11.38.269.2014 and 11.38.237.2015).


\newpage
\begin{longtable*}{l....}
\caption{ Individual contributions to the Lamb shift, in units of $F(Z\alpha)$. $Z$ is the nuclear
charge, $^1$H, $^4$He, $\ldots$ are the isotope symbols, $R$ is the root-mean-square nuclear charge
radius, $M/m$ is the ratio of the nuclear mass and the electron mass. Labelling of individual
contributions is explained in the text. The numbers in parentheses denote the uncertainty in the
last figure(s). The entries ``FNS'', ``SE(fns)'', ``Ue(fns)'', ``WK(fns)'' have two uncertainties;
the first one is due to the model-dependence, numerical and interpolation errors, whereas the
second one is due to the error of the nuclear charge radius. The total Lamb shift values have two
uncertainties; the second one is due to the error of the nuclear charge radius, whereas the first
one represents all other errors of individual contributions added quadratically. \label{tab:lamb} }
\\ \colrule\hline
  & \multicolumn{1}{c}{$1s_{1/2}$}  & \multicolumn{1}{c}{$2s_{1/2}$}  & \multicolumn{1}{c}{$2p_{1/2}$}  & \multicolumn{1}{c}{$2p_{3/2}$} \\
\hline\\[-5pt]
\multicolumn{5}{l}{
  $Z = 1,\ \mbox{\rm $^{1}$H },\  R = 0.8775\,(51)\ \mbox{\rm fm}, \  M/m =    1836.1527$}
 \\[2pt]
SE(pnt) & 10.316x794 & 10.546x825 & -0.126x396 & 0.123x499 \\
Ue(pnt) & -0.264x381 & -0.264x392 & -0.000x003 & -0.000x001 \\
WK(pnt) & 0.000x003 & 0.000x003 & 0.x & 0.x \\
FNS & 0.001x483\,(0)(17) & 0.001x483\,(0)(17) & 0.x & 0.x \\
\hline\hline
\end{longtable*}


\newpage

\begin{longtable*}{l....}
\caption{Ionization energy of the ground state and the energy level separation intervals in
hydrogenlike atoms, in Rydbergs. $Z$ is the nuclear charge, $^1$H, $^4$He, $\ldots$ are the isotope
symbols, $R$ is the root-mean-square nuclear charge radius in fermi, $M/m$ is the ratio of the
nuclear mass and the electron mass, ``DIRAC'' denotes the binding energy or the transition energy
according to the Dirac equation with the point nucleus $E_{\rm D}$, ``MASS'' denotes the leading
recoil correction $E_{\rm M}$,  ``LAMB'' denotes the Lamb shift $E_{\rm L}$, and ``TOTAL'' denotes
the total binding energy or transition energies. The numbers in parenthesis denote uncertainties in
the last figure(s). \label{tab:energy}}
\\ \colrule\hline
   &  \multicolumn{1}{c}{$1s$}  &  \multicolumn{1}{c}{$2s-1s$}  &  \multicolumn{
 1}{c}{$2p_{1/2}-2s$}  &  \multicolumn{1}{c}{$2p_{3/2}-2p_{1/2}$} \\
\hline\\[-5pt]
\multicolumn{5}{l}{
  $Z = 1,\ \mbox{\rm $^{1}$H },\  R = 0.8775\,(51)\ \mbox{\rm fm},\  M/m = 1836.1526724\,(8)$}
 \\[2pt]
DIRAC & -1.000013x3131931 & 0.750009x1528147 & 0x & 0.000003x328320432\,(2) \\
MASS & 0.000544x3205755\,(2) & -0.000408x2386190\,(2) & 0x & -0.000000x001812648 \\
LAMB & 0.000002x4842735\,(43) & -0.000002x1666270\,(43) & -0.000000x32154868\,(54) & 0.000000x007707557\,(33) \\
TOTAL & -0.999466x5083441\,(43) & 0.749598x7475687\,(43) & -0.000000x32154868\,(54) & 0.000003x334215341\,(33) \\
\hline\hline
\end{longtable*}


\newpage


\begin{table*}
\caption{Comparison with the previous compilation by Johnson and Soff \cite{johnson:85} for the
individual contributions to the Lamb shift,
in terms of $F(\Za)$.
``TW'' denotes the results of the this work, ``J\&S'' denotes the results of Ref.~\cite{johnson:85}. Labelling of individual
contributions is as in Ref.~\cite{johnson:85}.
\label{tab:js}
}
\begin{ruledtabular}
  \begin{tabular}{l........}
& \multicolumn{2}{c}{$1s$}
& \multicolumn{2}{c}{$2s$}
& \multicolumn{2}{c}{$2p_{1/2}$}
& \multicolumn{2}{c}{$2p_{3/2}$}
 \\\hline\\[-3pt]
& \mbox{\rm TW} & \mbox{\rm J\&S }
& \mbox{\rm TW} & \mbox{\rm J\&S }
&\mbox{\rm TW} & \mbox{\rm J\&S }
& \mbox{\rm TW} & \mbox{\rm J\&S }
 \\\hline\\[-3pt]
\multicolumn{9}{l}{
  $Z = 5$}
 \\[4pt]
 SELF & 6.251x\,627 & 6.251x\,6\,(10)       & 6.484x\,860 & 6.484x\,1\,(10)        & -0.122x\,775 & -0.122x\,6\,(2) & 0.125x\,623 & 0.125x\,6\,(2) \\
 S-FS & -0.000x\,010 &                      & -0.000x\,010 &                       & 0.x  & & 0.x & \\
 UEHL & -0.256x\,677 & -0.256x\,7           & -0.256x\,922 & -0.256x\,9            & -0.000x\,084 & -0.000x\,1 & -0.000x\,018 & 0.x \\
\end{tabular}
\end{ruledtabular}
\end{table*}


\begin{table*}
\caption{
Comparison of theory and experiment for H-like ions, in eV. 1~Ry = 13.605\,692\,53\,(30)~eV. $E_{\rm LM} = E_{\rm L} + E_{\rm M}$ is the sum of the Lamb shift and the
leading recoil correction, as defined by Eq.~(\ref{1a}).
\label{tab:comparison}
}
\begin{ruledtabular}
  \begin{tabular}{lccllcl}
$Z$ & Element &       & \multicolumn{1}{c}{Theory} & \multicolumn{3}{c}{Experiment} \\
& &       & & \multicolumn{1}{c}{Result} & Year & Ref. \\
\hline\\[-5pt]
2   &  $^4$He & $2p_{1/2}-2s$ & $0.000\,058\,070\,8\,(1)$ & $0.000\,058\,069\,4\,(7)$ & 2000 & \cite{wijngaarden:00} \\
%
3   & $^6$Li   & $2p_{1/2}-2s$ & 0.000\,259\,45\,(1) & 0.000\,259\,58\,(9)  & 1975 & \cite{leventhal:75} \\[2pt]
7   &$^{14}$N & $2p_{3/2}-2s$ & $0.103\,516\,8\,(1)$ & $0.103\,53\,(6)$  & 2004 & \cite{hosaka:04}\\
    &         &               &                       & $0.103\,519\,(10)$ & 2001 & \cite{myers:01}\\[2pt]
9   &$^{19}$F & $2p_{3/2}-2s$  &  $0.284\,748\,8\,(3)$& $0.284\,76\,(14)$ & 1975 & \cite{kugel:75} \\
\end{tabular}
\end{ruledtabular}
\end{table*}

\end{document}